\documentstyle[amssymb,aps]{revtex}
\input epsf
\def\erf{\mbox{erf}}
\def\At{{\tilde A}}
\def\Bt{{\tilde B}}
\def\Ac{{\cal A}}
\def\Bc{{\cal B}}
\def\Act{{\tilde {\cal A}}}
\def\Bct{{\tilde {\cal B}}}

\draft

\begin{document}

\title{Front dynamics during diffusion-limited corrosion
\\ of ramified electrodeposits.}

\author{Christophe L\'eger and Fran\c{c}oise Argoul \\
{\small {\it Centre de Recherche Paul Pascal, Avenue Schweitzer, 33600
Pessac, France}}
\\ Martin Z. Bazant \\
{\small {\it Department of Mathematics, Massachusetts Institute of
Technology, Cambridge, MA 02139}} \\
}
\date{\today}
\maketitle
\begin{abstract}

Experiments on the diffusion\--limited corrosion of porous copper
clusters in thin gap cells containing cupric chloride are reported.
By carefully comparing corrosion front velocities and 
concentration profiles obtained by phase-shift
interferometry with theoretical predictions, it is demonstrated that
this process is well-described by a one-dimensional mean-field model
for the generic reaction $\mbox{A} + \mbox{B\ (static)} \rightarrow
\mbox{C\ (inert)}$ with only diffusing reactant (cupric chloride) and
one static reactant (copper) reacting to produce an inert product
(cuprous chloride).  The interpretation of the experiments is aided by
a mathematical analysis of the model equations which allows the
reaction-order and the transference number of the diffusing species
to be inferred. Physical arguments are
given to explain the surprising relevance of the
one-dimensional mean-field model in spite of the complex (fractal)
structure of the copper clusters.

\end{abstract}

\section{Introduction}

Diffusion\--limited processes are ubiquitous in physics
\cite{Jacob:93}, chemistry \cite{Epstein:96} and
biology \cite{Jacob:97}. Reaction\--diffusion processes have been the
subject of intense and continuous interest since the work of
Smoluchowski \cite{Smoluchowski:16,Collins:49,Calef:83}.  A crucial
feature of many such processes controlling pattern formation and
reaction efficiency is the \lq\lq reaction front'', a dynamic but
localized region where reactions are most actively occurring which
separates regions rich in the individual reactants. The simplest
theoretical model of a reaction front, introduced more than a decade
ago by G\'alfi and R\'acz~\cite{Galfi:88}, is the \lq\lq mean-field''
model for two initially separated species A and B reacting to produce
an inert species C. Since then, the case of two diffusing reactants A
and B has been thoroughly studied
analytically~\cite{Galfi:88,Schenkel:93,Koza:96,Bazant:99} and
numerically~\cite{Jiang:90,Chopard:91,Cornell:91,Cornell:92,Cornell:93,Cornell:95,Cornell:97,Araujo:92,Araujo:93,Koza:96b,Taitelbaum:96b},
and some predictions of the mean-field model have been checked in
experiments~\cite{Taitelbaum:96b,Koo:90,Koo:91,Taitelbaum:92,Taitelbaum:96,Yen:96,Yen:97,Yen:97b,Yen:98,Taitelbaum:98}.

In contrast, the case of only one diffusing reactant A and one static
reactant B (confined on a fixed matrix) has not yet been studied
experimentally.  We show in this paper that the corrosion of a porous
solid (B) immersed in a chemically active fluid suspension (A) can
also be described by such a mean-field model.  Some
analytical~\cite{Bazant:99,Koza:97} and
numerical~\cite{Jiang:90,Havlin:95} studies exist for this case as
well, but since it is more microscopically complex (for a real porous
interface) than the case of two diffusing reactants (in a homogeneous
medium) an experimental test of the model is needed.

The mean-field model of a planar reaction front for the chemical
reaction
\begin{equation}
\mbox{A (diffusing) \ + \ B (static) \ $\rightarrow$ \ C
(inert)} ,
\end{equation}
postulates that the concentrations $\rho_A(X,T)$ and $\rho_B(X,T)$ of
species A  and B, respectively, evolve according to a
pair of coupled partial differential
equations~\cite{Bazant:99,Koza:97},
\begin{eqnarray}
\frac{\partial \rho_A}{\partial T} & = & D_A \frac{\partial^2
  \rho_A}{\partial X^2} - R(\rho_{A},\rho_{B}) \label{eq:rhoA} \\
\frac{\partial \rho_{B}}{\partial T} & = & - R(\rho_{A},\rho_{B}),
\label{eq:rhoB}
\end{eqnarray}
where $D_A$ is the diffusion constant for species A and
$R(\rho_A,\rho_B)$ is the reaction rate density.  The most frequently
used initial conditions assume that the reactants are uniformly
distributed and completely separated at first, $\rho_{A}(X,0) =
\rho_A^o H(X)$ and $\rho_B^o(X,0) =\rho_{B}^o H(-X)$, where $H(X)$ is
the Heaviside unit step function.  Such initial conditions are easier
to reproduce in experiments than those involving uniformly mixed
reactants.  There are several assumptions behind
Eqs.~(\ref{eq:rhoA})--(\ref{eq:rhoB}): $(i)$ The product $C$ is
generated in small enough quantities that its presence does not
significantly affect the dynamics; $(ii)$ The concentrations are
dilute enough that the diffusivities are constant;
$(iii)$ The fixed matrix containing reactant B
(static) is porous enough that reactant A can freely diffuse through it;
and $(iv)$ The reaction rate is a function of only the local
concentrations and not any fluctuations or many-body effects (which is
the \lq\lq mean-field approximation'').  It is common to make the
mean-field approximation under the assumption $R(\rho_A,\rho_B) = k
\rho_A^m \rho_B^n$, but in the interpretation of our experiments we
will not assume anything about the form of $R(\rho_A,\rho_B)$ {\it a
priori} since the reaction takes place at a solid\--liquid
interface. Moreover, this interface is highly ramified, and therefore,
the underlying microscopic dynamics is expected to be more complex
than for simple homogeneous kinetics.

In this paper we carefully test the validity of these assumptions with
experiments on a particular porous-solid corrosion system: copper
clusters corroded by a cupric chloride (CuCl$_2$) electrolyte. The
clusters are obtained by a thin\--gap cell electrodeposition from a
CuCl$_2$ electrolyte at fixed current. This process builds a depletion
layer of CuCl$_2$ ahead of the copper deposit. When the current is
switched off, this CuCl$_2$ depletion layer relaxes toward the copper
cluster bringing Cu$^{2+}$ cations which react with copper according
to:
\begin{equation}
\mbox{CuCl$_2$ (aq) \ + \ Cu (solid, red) \ $\rightarrow$ \
2 CuCl (solid, white)} \label{eq:bilan}
\end{equation}
where the cuprous chloride (CuCl) is produced in the form of small
(white) crystallites which drop down to the bottom of the cell.

In section II we describe the experimental set\--up and the method
used to prepare the porous clusters to be corroded. In section III we
report the experimental evidence that our corrosion system behaves
like a 1D diffusion\--reaction process with one static reactant. In
section IV, a mathematical analysis is presented which makes
quantitative predictions based on the experimental data of section II,
within the theoretical framework of the mean-field model,
Eqs.~(\ref{eq:rhoA})--(\ref{eq:rhoB}).  In the last section IV, the
experimental results are revisited to refine the comparison with the
theoretical model and to discuss in some detail of its physical
limitations.

\section{Experimental Methods}

\subsection{Apparatus}

\begin{figure}
\vspace*{1.5cm}
\epsfxsize= 12.0cm
\centerline{\epsfbox{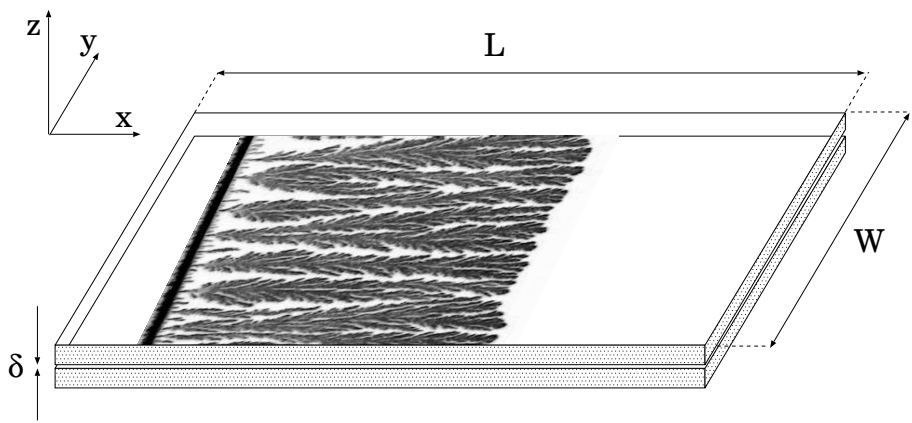}}
\vspace*{0.5cm}
\smallskip
  \caption{Schematic diagram of the thin-gap electrodeposition cell
containing a ramified, metallic copper deposit. Note that the size of
the deposit has been enlarged for clarity.}
\label{fig:cell}
\end{figure}

The experiments are performed in a thin-gap electrodeposition cell,
which is depicted schematically in Fig.~\ref{fig:cell}. The cell
consists of an unsupported, aqueous solution of CuCl$_2$ confined to a
narrow region of dimensions $W = 5 \mbox{cm} \times L = 8
\mbox{cm}\times \delta = 50\mu\mbox{m}$ between two closely spaced, optically
flat glass plates ($\lambda/4$ over 80mm$\times$50mm). Two parallel,
ultrapure copper and silver wires (50$\mu$m diameter, Goodfellow 99\%
purity) are inserted between the two glass plates to act both as
spacers and as electrodes. During the electrodeposition (prior to
corrosion) the wires are polarized so that the silver wire acts as the
cathode and the copper wire as the anode. The solutions of CuCl$_2$
(ACS reagent) are prepared from deionized water, carefully cleaned of
any trace of dissolved oxygen by bubbling nitrogen through it for one
hour. The anodic part of the cell (not shown in Fig.~\ref{fig:cell})
is filled by a dilute solution of CuCl$_2$ to postpone the
precipitation of the salt due to saturation effects by dissolution of
the anode. The copper electrodeposits are all grown at constant
current, and the entire experiment is performed at room temperature
($\approx 20^{\circ}C$).

Digitized color pictures of the copper clusters are obtained by direct
imaging of the cluster through a lens, using a three\--CDD camera
coupled with an $8$-bit frame grabber from Data Translation driven by
the public domain software IMAGE \cite{NIH} which successively
captures three RGB frames and from them reconstructs the color image.

A phase-shift Mach Zehnder interferometer is used independently to
resolve the concentration field, averaged over the depth of the cell.
A sketch of the interferometer can be found in ref.\cite{Leger:97}.
The interference patterns are recorded through a CCD 
camera coupled to the same frame grabber \cite{NIH} with a
768$\times$512 pixel resolution. Phase-shift interferometry offers
several significant advantages over traditional interferometry in that
it provides an accurate reconstruction of the entire concentration
field, using a set of successive interference pictures recorded for
shifted values of the phase difference between two optical wavefronts,
and can also be used as an holographic
interferometer~\cite{Robinson:93}.

\subsection{Preparation of Copper Clusters by Electrodeposition}

When current flows from the anode to the cathode, charge transfer
occurs at the cathode, leading to the reduction of copper cations into
copper metal according to~\cite{Leger:97,Leger:98,Oberholtzer:98}:
\[
\mbox{Cu}^{2+} + 2 \mbox{e}^- \rightarrow \mbox{Cu} \; \; .
\]
The actual mechanism of deposition is much more complex than this 
two electron transfer process since competitive reactions involving the
solvent species are likely to occur.
Nevertheless, in CuCl$_2$
electrolytes, we have observed that the formation of cuprous oxide (Cu$_2$O)
in competition with copper by reduction of Cu$^{2+}$ cations is not
favored, contrary to what is observed in copper sulphate (CuSO$_4$)
solutions~\cite{Trigueros:94,Lopez:96,Texier:98},
which can be partly explained by the strong adsorption and
complexation properties of chloride anions~\cite{Oberholtzer:98}.
This reduction process on the cathode
implies a local depletion of the copper cations close to the cathode
and, therefore, also their replenishment by a global transport
process, namely diffusion. Although electromigration also contributes
to transport, it does not act independently of diffusion in regions
where electroneutrality is maintained~\cite{Newman:91}, which 
means everywhere in the cell outside the $10-100\AA$ thick double
layer~\cite{Parsons:80,Bard:80}. This often misunderstood fact was
given a firm theoretical basis by Newman over 30 years ago in his
asymptotic analysis of the transport equations for a rotating disk
electrode~\cite{Newman:65}, but only recently has it been
quantitatively verified in experiments (by our group) for the case of
constant boundary flux at a fixed cathode~\cite{Leger:97,Leger:98}. In
summary, the theoretical and experimental evidences indicate that in
the absence of convection the concentration $\rho_A$ of a dilute,
binary electrolyte evolves according to the classical diffusion
equation,
\begin{equation}
\frac{\partial\rho_A}{\partial t} = D_A \nabla^2 \rho_A,
\label{eq:diff}
\end{equation}
where $D_A$ is the \lq\lq ambipolar diffusion coefficient" for the
electrolyte given by a certain weighted average of the diffusion
constants of the individual ions~\cite{Newman:91}.

When the interfacial concentration of metal cation Cu$^{2+}$
approaches zero, the interface becomes unstable
and develops into a forest of fine spikes which compete between each
other to invade the cell~\cite{Chazalviel:90,Elezgaray:98}. In some cases, a
``dense-branched'' pattern is selected
\cite{Leger:98,Melrose:90,Fleury:90,Fleury:91,Bazant:95}. This
morphology is characterized by a dense array of branches of invariant
width advancing at constant velocity $v$ through the cell, whose tips
delimit a nearly planar front between the copper
salt electrolyte and the deposit zone. We have shown recently that
this growth regime can be modeled {\it via} a 1D diffusion model through the
measurement of the copper salt concentration field ahead of the
growing deposit by interferometry \cite{Leger:98,Bazant:95}. The experimental
concentration field closely fits the \lq\lq traveling-wave" solution
to Eq.~(\ref{eq:diff}),
\begin{equation}
\rho_A(X) = \rho_A^o \left( 1-\exp ^{-X/L_d} \right)
\label{eq:expeini}
\end{equation}
where $L_d=D_A/v$, and $X$ is the distance to the front edge of the
copper deposit, in the direction normal to the front, oriented toward
the bulk electrolyte. The diffusion length $L_d$ is proportional to
$\rho_A^o/j$, where $\rho_A^o$ is the initial bulk concentration in
copper cations and $j$ is the current density. This diffusion length
tends to zero as $j/\rho_A^o$ increases, and in that limit the
concentration profile looks like a step function. Note that
$\rho_A(X \leq 0)\approx 0$ and $\rho_A(X \gg L_d) \approx \rho_A^o$, i.e.
the metallic copper deposit leaves behind it a region entirely depleted in
copper cations pushing in front of it a diffusion layer of constant width
extending into the bulk electrolyte. Due to the conservation of copper
during the deposition process, a linear relation exists between the
velocity $v$ of the growth and the interfacial flux of cations $J$,
namely $v \rho_B = J$, where $\rho_B$ is the mean concentration of
(metallic) copper in the region of the deposit.

Using the relation $v=D_A/L_d$, the ratio of the copper concentration
in the bulk electrolyte $\rho_A^o$ (where it takes the form of cupric
ions) to that in the region of the deposit $\rho_B^o$ (where it mostly
takes the metallic form) is easily calculated from the basic properties of
the electrolyte~\cite{Leger:98,Fleury:90,Fleury:91,Bazant:95}
\begin{equation}
q \ \equiv \  \frac{\rho_A^o}{\rho_B^o} \ = \  1-t^+ , \label{eq:q}
\end{equation}
where $t^+$ is the transference number \cite{Newman:91,Bard:80} of the
copper cation in a CuCl$_2$ electrolyte.  Practically, 
$t^+$ is a characteristic of the electrolyte and therefore $q$ 
will not be a free parameter in our experiments (neither
$t^+$ nor $q$ depend on the current density $j$).  The closer $t^+$ is
to 1, the greater the concentration of copper inside the cluster.  In
CuCl$_2$ electrolytes, $t^+$ is expected to be smaller than 0.5, which
implies that the copper composition of the deposited zone will not go
beyond twice the original concentration of CuCl$_2$ in the
electrolyte. Therefore, the copper clusters obtained by thin gap
electrodeposition in CuCl$_2$ are in fact highly porous.

The large porosity of the deposited copper
clusters is of fundamental importance in our subsequent study of the
corrosion of the copper deposits once the current has been switched
off (and the electrodeposition halted) because, as a consequence, the
cupric ions are able to diffuse freely through the dendrites with
approximately their bulk diffusivity and then react with a large
exposed surface of metallic copper. The low density of the deposit
also suggests that the product of the corrosion reaction, cuprous
chloride (CuCl) crystal, is produced in small enough quantities that
its presence should not significantly affect the dynamics of the
reaction-diffusion process. Therefore, by interrupting the current
during electrodeposition we can observe a simple reaction-diffusion
system with two initially separated reactants, copper chloride (A)
and metallic copper (B), only one of which is free to diffuse. Since
the initial interface between the bulk electrolyte and the ramified
electrodeposit is planar and the deposit is disordered, it is likely
that the dynamics of the corrosion process will be effectively \lq\lq
one-dimensional" (1D), in the sense that there might be nearly perfect
translational symmetry in the two spatial directions ($Y$ and $Z$)
perpendicular to direction of the front propagation. Moreover, since
the dynamics occurs in three dimensions (as opposed to two for a
surface or one for a molecular channel),
it is also likely that a mean-field, continuum model will be valid,
although this may not seem obvious {\it a priori} in light of the
complex geometry of the electrodeposits, which is known to be fractal
~\cite{Kuhn:94,Kuhn:95,Argoul:97}.

The rest of the paper is devoted to a careful, experimental validation
of these hypotheses, showing that our system is indeed accurately
described by a one-dimensional, mean-field model for the generic
chemical reaction, A$ + $B$\mbox{(static)} \rightarrow $C. We begin in
the next section by describing the scaling behavior of the reaction
front and accompanying depletion layer of CuCl$_2$. In the following
section, a mathematical analysis of the one-dimensional, mean-field
model is presented which incorporates the observed scalings and makes
quantitative predictions regarding the reaction front speed and the
concentration evolution. Finally, these predictions are checked with a
more detailed analysis of the experimental data in the last section,
and arguments are given to explain the relevance of the
one-dimensional, mean-field model for our experimental system.

\section{Preliminary Experimental Results}

\subsection{Temporal Evolution of the Corrosion Front}

\begin{figure}
\vspace*{1.5cm}
\epsfxsize= 12.0cm
\centerline{\epsfbox{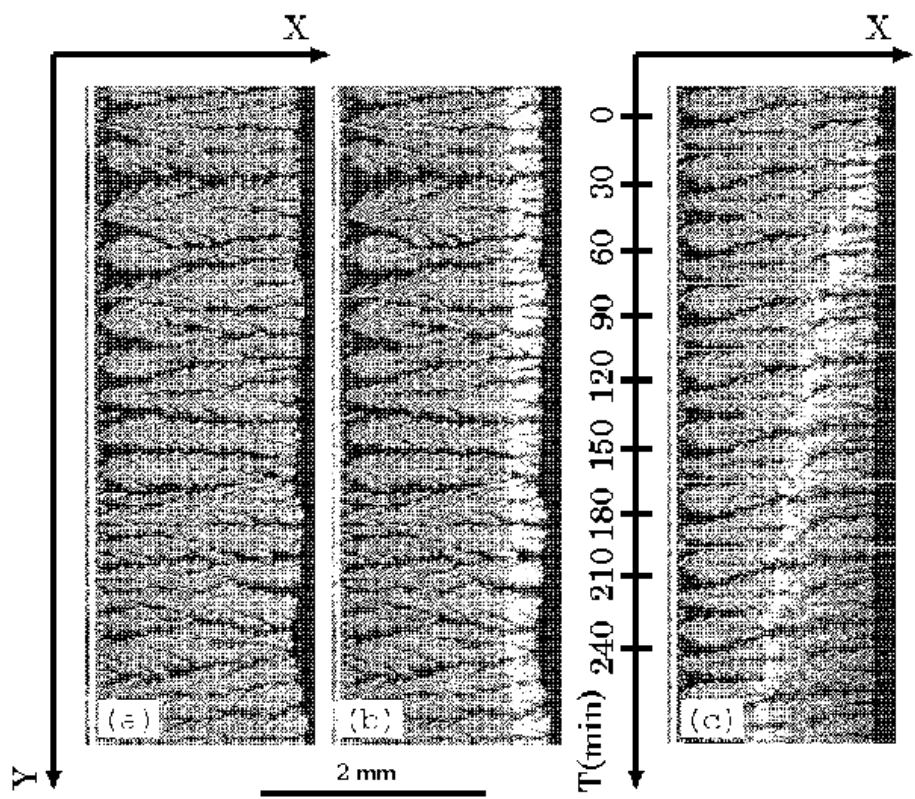}}
\vspace*{0.5cm}
\smallskip
  \caption{(a) Photograph of a copper deposit grown from a
    0.5~mol.l$^{-1}$ CuCl$_2$
    solution at $j$~=~40~mA.cm$^{-2}$ for approximately 15 minutes.
(b) Photograph of the same deposit half an hour after the current had
been switched off. (The white zone is CuCl.) (c) The montage shows a
sequence of photographs of a small region of the deposit including the
reaction front taken every 30 minutes after the interruption of the
current.} 
\label{fig:montage}
\end{figure}

At the moment when the current is switched off, the region of the
copper deposit is entirely depleted of cupric ions, which are thus
initially separated from the metallic copper in the deposit. At later
times, cupric ions diffuse amidst the copper dendrites and react at
the metal surfaces, leaving behind cuprous chloride (CuCl) crystallites. In
Fig.~\ref{fig:montage} (a) and (b) are shown images of a copper
deposit just prior to corrosion and after 30 minutes of corrosion,
respectively. Note that in Fig.~\ref{fig:montage} (b) the interfacial
region between the red copper (the grey color in this picture) and the
white CuCl is rather flat and thin.

Focusing on the temporal evolution of this red/white interface, we
have observed that, while at first the white layer of CuCl
appears at the tips of the copper-deposit branches, it gradually
becomes flatter and flatter. As a result, the system approaches
translational invariance along the $Y$ direction, normal to the growth
direction $X$, thus justifying a one-dimensional model for the system
involving the single spatial coordinate $X$ (normal to the reaction
front).

By carefully comparing the concentration field of cupric ions obtained
by phase-shift interferometry and the red/white,
Cu/CuCl interface observed on the deposit, the
location and extent of the reaction front, where there is a
significant overlap of metallic copper and cupric ions, can be
identified. Following a transient regime (which we describe in the
last section), it is observed that the reaction front approaches a
constant width $w \sim T^\alpha$ with $\alpha = 0$, which is
consistent with certain mean-field
models~\cite{Bazant:99,Jiang:90,Koza:97}. Using the theoretical
methods pioneered by G\'alfi and R\'acz~\cite{Galfi:88} in the
case of two diffusing reactants, this scaling was first predicted by
Jiang and Ebner~\cite{Jiang:90} using physical arguments supported by
computer simulations and later by Koza~\cite{Koza:97} using asymptotic
analysis. 

Recently, Bazant and Stone~\cite{Bazant:99} have considered the case
of higher-order reactions $mA + nB\mbox{(static)}
\rightarrow C$ represented by the mean-field reaction rate
$R(\rho_A,\rho_B) = k \rho_A^m\rho_B^n$ and proved that the scaling
exponent for the front width is given (uniquely) by the
formula
\begin{equation}
\alpha = \frac{m-1}{2(m+1)} ,
\end{equation}
which holds for any real number $m \geq 1$. (The scaling solution does
not exist for $m < 1$.) In light of this result, the experimental
observation $\alpha = 0$ is consistent with the usual one-dimensional,
mean-field theory only in the case $m=1$. If higher-order reactions
were present $m > 1$, the theory would predict that the reaction front
width increases in time ($\alpha > 0$) although always more slowly
than diffusion ($\alpha < \frac{1}{2}$). 

The position of the reaction front $X_f(T)$ during the corrosion of a
copper deposit (grown from a 0.5M CuCl$_2$ solution at $j = 40$
mA/cm$^2$) is plotted in Fig~\ref{fig:front}. Note that, after initial
transients have vanished ($T > 500~s$), the reaction front itself
\lq\lq diffuses" with its position given by the scaling, $X_f \sim
T^\sigma$ with $\sigma = \frac{1}{2}$, which is is also consistent
with predictions of the one-dimensional mean-field
model\cite{Galfi:88,Jiang:90,Koza:97}. In fact, this diffusive
movement of the reaction front after long times is a robust feature of
all mean-field models for two initially separated species, regardless
of the reaction orders $m, n \geq 1$ or the number of diffusing
reactants (one or two), as long as there is no relative advection of
the two species (due to fluid flow or some other external forcing)
\cite{Bazant:99} or impermeable membrane to one of the reactants
\cite{Chopard:97}.

\subsection{Temporal Evolution of the Diffusion Layer}

At $T = 0$ when the current is interrupted, the reactants Cu and
CuCl$_2$ are completely separated, since the concentration of CuCl$_2$
is negligibly small in the immediate vicinity of the metallic Cu
electrodeposit. During the subsequent corrosion process the
concentration of CuCl$_2$ remains very small in the reaction front,
which leads to the modification of the initial depletion layer of
CuCl$_2$ (produced by the electrodeposition process) into a region where
the concentration smoothly interpolates to the value of the bulk
solution far behind the front. The term \lq\lq diffusion layer" is
used to describe this region because it is characterized by the
transport of fresh CuCl$_2$ by diffusion from the bulk, relatively
unaffected by chemical reactions due to the negligible (or vanishing)
concentration of metallic Cu remaining behind the reaction front.

\begin{figure}
\vspace*{1.5cm}
\epsfxsize= 9.0cm
\centerline{\epsfbox{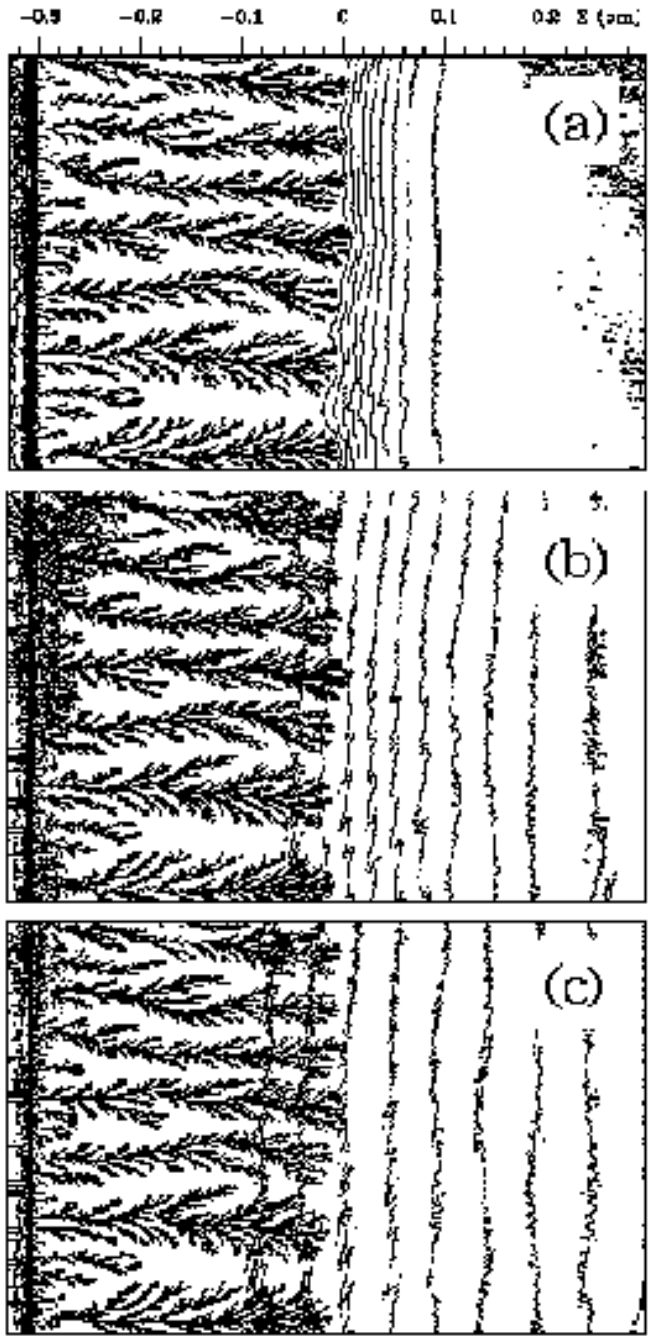}}
\vspace*{0.5cm}
\smallskip
  \caption{Interferometric characterization of the concentration field
  around a copper deposit during its dissolution (a) just before the
  interruption of the current, (b) 15 minutes and (c) one hour
  later. ($\Delta \rho \approx \rho_A^o/10$ between adjacent
  isoconcentration lines.) The deposit grown in 0.5~mol.l$^{-1}$ CuCl$_2$
  solution at $j=40$~mA.cm$^{-2}$ for 20 minutes. 
  The concentration of CuCl$_2$ is negligibly small inside and
  ahead (to the left) of the reaction front and approaches the bulk
  value of 0.5~mol.l$^{-1}$ far behind (to the right of) the front. }
\label{fig:pattern}
\end{figure}

The temporal evolution of the diffusion layer is revealed by precise
interferometric measurements of the concentration profile of CuCl$_2$.
In Fig.~\ref{fig:pattern} are shown three contour plots of the
concentration field computed from the integrated index along the depth
of the cell. Since the experiments are performed in thin-gap cells
($50 \mu$m) and the depletion layer spreads over distances larger than
this gap, it is safely assumed that the concentration of CuCl$_2$ does
not change appreciably along the $Z$ direction (parallel to the laser
beam \cite{Leger:98}). 
If Fig.~\ref{fig:pattern}, the shadow of the Cu/CuCl cluster is
also clearly seen. A close inspection of the panels (b) and (c), which
correspond to eroded clusters, reveals that in the zone of the copper
cluster where CuCl$_2$ has diffused (recognizable where the leftmost
isoconcentration contours have moved through the cluster), the cluster
has been broken down in smaller crystallites, which, as indicated by
their color in Fig.~\ref{fig:montage}, are made of CuCl.

Typical experimental concentration profiles of CuCl$_2$ measured at
different times (averaged along the $Y$ direction, normal to the
growth direction) are shown in Fig.~\ref{fig:profils1}. The shape of
these concentration profiles is discussed in the next two sections,
but here we focus on the scaling of the width $W_d(T)$ of the
diffusion layer (defined as the region of non-negligible gradients).
Fig.~\ref{fig:gradient} shows that at long times ($T > 500~s$), the
diffusion layer approaches a self\--similar structure, with the
diffusive flux entering the reaction front obeying the scaling law,
$J_d \propto (\partial \rho_A/\partial X)|_{X=X_f} \sim T^{-\delta}$,
and that, therefore, the width of the diffusion layer has the familiar
scaling~\cite{Galfi:88,Koza:96,Jiang:90,Koza:97}  $W_d \sim
T^\delta$ with $\delta = \frac{1}{2}$, which is another robust
feature of the mean-field models~\cite{Bazant:99}.

A physical argument based on mass conservation between the diffusion
layer and reaction front~\cite{Galfi:88,Jiang:90} can be used to
predict the scaling of the reaction rate (per unit volume) in the front
$R \sim T^{-\beta}$ from the preceding experimental
observations. The total reaction rate in the front (per unit area)
scales as $w R
\sim T^{\alpha-\beta}$, and this flux of cupric ions due to reactions
must balance the diffusive flux entering the front $J_d
\sim T^{-\delta}$, which yields the scaling relation, $\beta =
\alpha + \delta = 0 + \frac{1}{2} = \frac{1}{2}$. It is important to
point out, however, that while $\alpha = 0$ and $\delta=\frac{1}{2}$
are the results of direct experimental observations, the scaling
exponent $\beta = \frac{1}{2}$ is only inferred by a physical
argument, based on the assumption that chemical reactions are
negligible in the diffusion layer. Although this assumption has been
checked numerically and analytically for various mean-field models,
the reaction rate is not directly measured in our experiments.

In the general case $R(\rho_A,\rho_B) = k \rho_A^m\rho_B^n$ mentioned
above, it can be shown~\cite{Bazant:99} that $\beta$ is given (uniquely) by
\begin{equation}
\beta = \frac{m}{m+1} ,
\end{equation}
so that once again $m=1$ is suggested by the inferred value $\beta =
\frac{1}{2}$. However, given that the experimental system has complex
fractal structure and three-dimensional transport in the reaction
front, it is not obvious {\it a priori} that $R(\rho_A,\rho_B) = k
\rho_A^m\rho_B^n$ is a reasonable approximation within a
spatially averaged one-dimensional model. Instead, we will make no
{\it ad hoc} assumptions about the functional form of the reaction
rate $R(\rho_A,\rho_B)$ and then explore consequences of only our
direct experimental observations, $\alpha=0$ and $\delta = \sigma =
\frac{1}{2}$, within the framework of a one-dimensional
mean-field model.

\section{Theoretical Predictions of the Mean-Field Model}

\subsection{Dimensionless Model Equations}

The model
equations have a dimensionless form involving only the parameter, $q
\equiv
\rho_A^o/\rho_B^o$, defined in Eq.~(\ref{eq:q}),
\begin{eqnarray}
\frac{\partial a}{\partial t} & = & \frac{\partial^2 a}{\partial x^2} -
r(a,b) , \label{eq:a} \\
\frac{\partial b}{\partial t} & = & - q\  r(a,b) , \label{eq:b}
\end{eqnarray}
with boundary and initial conditions
\begin{eqnarray}
a(-\infty,t) = 0, & \ \ \ & a(\infty,t) = 1 \\
b(-\infty,t) = 1, & \ \ \ & b(\infty,t) = 0 , \label{eq:bc} \\ 
a(x,0) = H(x), & \ \ \ & b(x,0) = H(-x) 
\label{eq:init}
\end{eqnarray}
where 
\begin{equation}
a \equiv \frac{\rho_A}{\rho_A^o} , \ \  \  b \equiv
\frac{\rho_B}{\rho_B^o}, 
\end{equation}
\begin{equation}
r(a,b) \equiv \frac{R(a\rho_A^o,b\rho_B^o)}{R(\rho_A^o,\rho_B^o)} 
\end{equation}
\begin{equation}
t \equiv \frac{R(\rho_A^o,\rho_B^o) T}{\rho_A^o}, \ \ \ 
x \equiv X \sqrt{\frac{R(\rho_A^o,\rho_B^o)}{(D_A\rho_A^o)}} .
\end{equation}
These initial conditions are closest to the actual ones used in the
experiments when the copper deposit is grown at large current, which
corresponds to small $L_d$ in Eq.~(\ref{eq:expeini}).  The
initial-boundary-value problem of Eqs.~(\ref{eq:a})--(\ref{eq:init})
involves an idealized, infinite system possessing no natural length or
time scale, and, therefore, it is expected that asymptotic similarity
solutions exist in which distance and time appear coupled by power-law
scalings~\cite{Barenblatt:79}. The experimental system, on the other
hand, possesses several relevant length scales, but they turn out not
to affect the evolution of the reaction front, at least for some range
of times. For example, the spatial scales of the copper deposit, such
as the typical dendrite spacing and dendrite width, surely affect the
dynamics at early times since these length scales are of the same
order as the diffusion length $L_d$ \cite{Leger:98}, but it is
observed that during corrosion the system quickly approaches planar
symmetry, averaged across scales much larger than individual
dendrites. Likewise the length scale of the gap spacing is not
expected to greatly influence the corrosion dynamics because vertical
(buoyancy-driven) convection, which has been observed during the
growth phase \cite{Huth:95} is suppressed in 50~$\mu$m depth cells
\cite{Leger:97,Chazalviel:96}.  However, the settling of the reaction
product, CuCl crystallites, could have some effect on the front
dynamics at this scale.  Finally, the largest length scales, namely
the distances from the outer edge of the deposit to the two
electrodes, also should not affect corrosion dynamics until the
reaction front gets close to the cathode and/or the diffusion layer
approaches the anode. Therefore, during intermediate times, after
three-dimensional transient effects have subsided but before the
system size begins to matter, the corrosion dynamics should be well
described by a self-similar solution to the one-dimensional mean-field
equations.

\subsection{The Diffusion Layer}

Motivated by these arguments and the experimental data, we consider
the transformation
\begin{equation}
a(x,t) = \At(\zeta,t), \ \ \ 
b(x,t) = \Bt(\zeta,t), \ \ \ 
\mbox{where} \ \ \zeta = \frac{x - x_f(t)}{2 \sqrt{t}},
\label{eq:Adef}
\end{equation}
for the concentration of CuCl$_2$ in the diffusion layer (defined by
$\zeta> 0$), and seek an asymptotic similarity solution, $\At(\zeta,t)
\sim A(\zeta)$ and $\Bt(\zeta,t) \sim B(\zeta)$ with power-law
expressions for $x_f(t)$ and $W_d(t)$. The experimental observations
discussed in the previous section support the scaling law $W_d \sim
t^{1/2}$ for the diffusion layer width and a similar diffusive scaling
law for the reaction front position $x_f
\sim t^{1/2}$. Therefore, we make the definitions
\begin{equation}
x_f(t)  = -2\nu\sqrt{t}\; , \label{eq:defxf} 
\end{equation}
\begin{equation}
\zeta  =  \frac{x}{2\sqrt{t}} + \nu \; , \label{eq:zeta}
\end{equation}
where $\nu(q)^2$ is an effective diffusion constant for the
reaction front to be determined during the analysis.

Substituting these expressions into Eq.~(\ref{eq:a}), we have,
\begin{equation}
\frac{\partial \At}{\partial t} + \left( \frac{\nu - \zeta}{2t} \right)
\frac{\partial \At}{\partial \zeta} = \left(\frac{1}{4t}\right)
\frac{\partial^2 \At}{\partial \zeta^2} - r(\At,\Bt) , \label{eq:Aall}
\end{equation}
which simply amounts to a change of variables from $(x,t)$ to
$(\zeta,t)$. We now look for an asymptotic similarity solution by
assuming that the time derivative vanishes relative to the diffusive
term,
\begin{equation}
\lim_{t\rightarrow\infty}\ t\ \frac{\partial \At}{\partial t} = 0, \ \
\mbox{for} \ \zeta > 0 \ \mbox{fixed} , \label{eq:Aqs}
\end{equation}
which has been called the \lq\lq quasistationary approximation" in the
physics literature~\cite{Koza:96,Cornell:93,Koza:97,Ben-Naim:92}. 
This is not really an \lq\lq approximation" but rather is an exact
asymptotic property of a certain class of self-similar solutions to
Eqs.~(\ref{eq:a})--(\ref{eq:bc}) which happen to accurately fit the
experimental data, as we will show in the next section. At this point
it is common to assume that the reaction term also vanishes relative
to the diffusion term,
\begin{equation}
\lim_{t\rightarrow\infty}\ t \cdot r(\At,\Bt) = 0, \ \
\mbox{for} \ \zeta > 0 \ \mbox{fixed} , \label{eq:Rzero}
\end{equation}
and that the concentration of the non-diffusing species also vanishes
in the diffusion layer, i.e. where the reaction front has already
passed,
\begin{equation}
\lim_{t\rightarrow\infty}\ \Bt(\zeta;t) = 0, \ \
\mbox{for} \ \zeta > 0 \ \mbox{fixed} . \label{eq:Bzero}
\end{equation}
Note that $\Bt(\zeta,t) = 0$ for $\zeta \geq \nu$ at all times due to
the initial condition $b(x,0) = H(-x)$ and the fact that this reactant
does not diffuse.  The limits in Eqs.~(\ref{eq:Rzero}) and
(\ref{eq:Bzero}) have previously been taken as {\it ad hoc}
assumptions~\cite{Koza:97}, but it can be shown that they are in fact
necessary consequences of the quasistationarity~\cite{Bazant:99}.

\begin{figure}
\vspace*{1.5cm}
\epsfxsize= 12.0cm
\centerline{\epsfbox{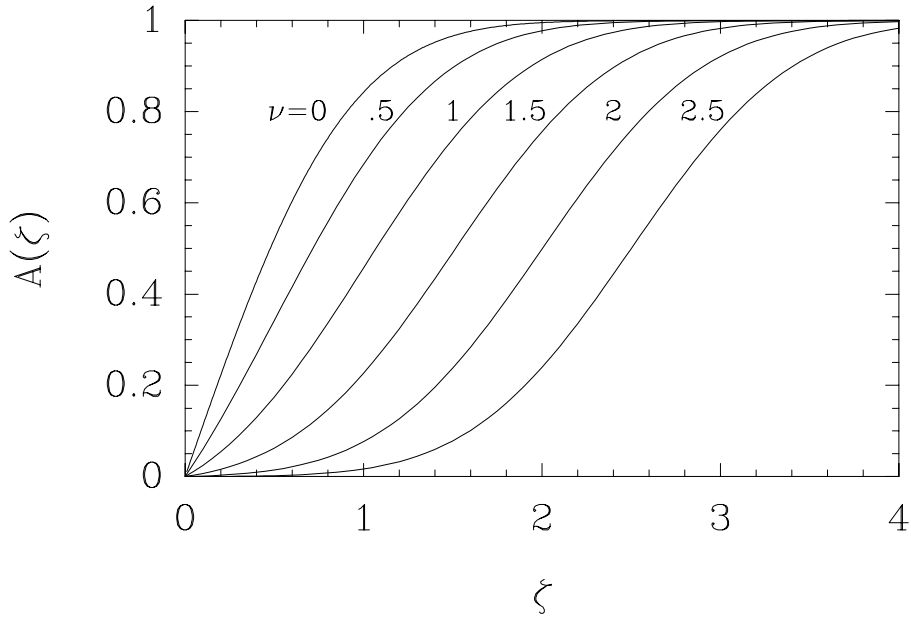}}
\vspace*{0.5cm}
\smallskip
  \caption{The asymptotic similarity function $a(x,t) \sim A(\zeta)$
    where  $\zeta=\frac{x}{2\protect\sqrt{t}}+\nu$ shown for $A_o=0$ and
    $\nu=$ 0, 0.5, 1, 1.5, 2, 2.5 from left to right.}
\label{fig:anu}
\end{figure}

Using Eqs.~(\ref{eq:Aqs})--(\ref{eq:Rzero}) and passing to the limit
$t \rightarrow \infty$ with $\zeta>0$ fixed in Eq.~(\ref{eq:Aall})
yields an ordinary differential equation for the asymptotic
diffusion-layer concentration $A(\zeta)$,
\begin{equation}
A^{\prime\prime} + 2(\zeta - \nu)A^\prime = 0 .
\label{eq:diffeq}
\end{equation}
The solution to this equation subject to the boundary condition
$A(\infty) = 1$ can be written in terms of error
functions~\cite{Abramowitz:65},
\begin{equation}
A(\zeta) = A_o + (1-A_o)\frac{\erf(\zeta-\nu) + \erf(\nu)}{1 +
\erf(\nu)} ,
\label{eq:diffA}
\end{equation}
where $A_o \equiv A(0)$ is a constant to be determined by asymptotic
matching with the reaction front as $\zeta \rightarrow 0$. The
function $A(\zeta)$ is shown in Fig.~\ref{fig:anu} for different
values of $\nu$. The slope of $A(\zeta)$ at $\zeta = 0$ given by
\begin{equation}
A^\prime(0) = \frac{2 (1-A_o) e^{-\nu^2}} {\sqrt{\pi}(1 +
\erf(\nu))} , \label{eq:A1}
\end{equation}
is the (dimensionless) diffusive flux into the reaction front.

On the length scale $W_d(t) \sim t^{1/2}$ appropriate for the diffusion
layer, the self-similar asymptotic concentration fields just derived
appear not to be differentiable at $\zeta = 0$,
\begin{equation}
a(x,t) \sim A(\zeta)H(\zeta), \ \ b(x,t) \sim H(-\zeta) , \ \
\mbox{as} \ t \rightarrow \infty \ \mbox{with} \ \zeta
\neq 0 \ \mbox{fixed},
\label{eq:AB}
\end{equation}
but, as we have already observed experimentally, that is only because
in reaction front (at $\zeta=0$) the concentrations are smoothly
interpolated across these apparent discontinuities on a much smaller
length scale $w \sim t^\alpha = o(W_d)$ since $\alpha < \delta$. In
mathematical terms, the asymptotic approximations in Eq.~(\ref{eq:AB})
are not uniformly valid for all $(x,t)$ as $t \rightarrow \infty$, but
rather are valid only for $\zeta\neq 0$, {\it i.e.}  $\sqrt{t} = O(|x
+ 2\nu\sqrt{t}|)$.

\subsection{The Reaction Front}

We now explore the consequences of the experimental results $\alpha =
0$ and $\delta = \sigma = \frac{1}{2}$ within the present mathematical
model. Although the physical arguments made above for the lack of a
natural length scale are much more tenuous in the reaction front
because the observed front width (about 0.2~mm) is comparable to the
average dendrite thickness (0.1~mm) and spacing (0.4~mm) as well as
the gap (0.05~mm), the nearly perfect planar symmetry of corrosion
process leads us to nevertheless seek another asymptotic similarity
solution to the one-dimensional, mean-field equations in the vicinity
of the reaction front, $x - x_f(t) = O(1)$.  The predictions of the
model will be carefully tested against the experimental data in the
next section.

Since $\alpha=0$ and $\sigma=\frac{1}{2}$, we consider the
transformation
\begin{equation}
a(x,t) = t^{-\gamma}\Act(\eta,t), \ \ b(x,t) = \Bct(\eta,t),
\label{eq:Acdef}
\end{equation}
where $\eta$ is a new similarity variable for the reaction front
defined by
\begin{equation}
\eta = x + 2\nu t^{1/2} = 2 t^{1/2} \zeta . \label{eq:etadef}
\end{equation}
The exponent $\gamma \geq 0$ is introduced to allow for the
possibility that $a(x,t) \rightarrow 0$ in the reaction front, which
is suggested by the result $r(a,b) \sim t^{-\beta}$ with $\beta =
\frac{1}{2}$ inferred earlier from the experimental data. In
contrast, no such prefactor multiplies $\Bct(\eta,t)$ in the reaction
front since $b(x,t)$ must remain finite there in order to interpolate
between the limiting values of $0$ and $1$, respectively, behind and
ahead of the front.

Making these transformations in Eq.~(\ref{eq:a}) yields
\begin{equation}
\frac{\partial \Act}{\partial t} + \nu t^{-1/2}
\frac{\partial \Act}{\partial \eta} - \gamma t^{-1} \Act =
\frac{\partial^2 \Act}{\partial \eta^2} - t^\gamma \
r\left(t^{-\gamma}\Act, \Bct \right) . \label{eq:Acall}
\end{equation}
As before, we explore the possibility of self-similar
quasistationarity in the reaction front: $\Act(\eta,t) \sim \Ac(\eta)$
and $\Bct(\eta,t) \sim \Bc(\eta)$ as $t \rightarrow \infty$ with
$|\eta| < \infty$ fixed. The consequence of the quasistationarity
assumption in Eq.~(\ref{eq:Acall}) is
\begin{equation}
\Ac^{\prime\prime}(\eta) = \lim_{t \rightarrow \infty}
\ t^{\gamma}\ r\left(t^{-\gamma}\Ac(\eta), \Bc(\eta) \right) \ \
, \ \ \mbox{for fixed} \; \;  \eta
\label{eq:rlimit}
\end{equation}
Since $\Ac^{\prime\prime}(\eta) = 0$ cannot satisfy the boundary
condition $\Ac(-\infty) = 0$ (except in the trivial case
$\Ac(\eta)=0$), the limit on the right-hand side of
Eq.~(\ref{eq:rlimit}) must be nonzero (and finite),
which is possible only if $r(a,b)$ is linear in $a$, i.e.
\begin{equation}
r(a,b) = a f(b) , \label{eq:r1}
\end{equation}
for some function $f(b)$. Therefore, the experimental facts $w(t)
\sim t^0$ and $x_f(t) \sim t^{1/2}$ are consistent with the
one-dimensional, mean-field model only if the reaction rate is first
order in the diffusing species.

Next we make the same transformation in Eq.~(\ref{eq:b}) and replace
the reaction term with Eq.~(\ref{eq:r1}) to obtain
\begin{equation}
\frac{\partial \Bct}{\partial t} + \nu t^{-1/2} \frac{\partial
\Bct}{\partial \eta} = - q t^{-\gamma}\Act f(\Bct) .
\end{equation}
By inspection, quasistationarity is possible only if $\gamma =
\frac{1}{2}$, which would imply $r(a,b) \sim t^{-1/2} \Ac(\eta)
f\left(\Bc(\eta)\right)$. Therefore, we conclude $\beta =
\frac{1}{2}$ once again, and the physical argument given in the previous
section is found to have sound mathematical justification.

With these results we arrive at a third-order system of
nonlinear ordinary differential equations for the concentration fields
in the reaction front,
\begin{eqnarray}
\Ac^{\prime\prime} - \Ac f(\Bc) & = & 0 , \\
\nu \Bc^{\prime} + q \Ac f(\Bc) & = & 0 .
\end{eqnarray}
These equations may be combined to eliminate the reaction term and
integrated once using the boundary conditions ahead of the front,
$\Ac(-\infty)=0$ and $\Bc(-\infty)=1$ to obtain
\begin{equation}
q \Ac^\prime = \nu (1 - \Bc) . \label{eq:Acint}
\end{equation}
Before proceeding with another integration, however, a third boundary
condition is needed, which comes from asymptotic matching with the
diffusion layer.

\subsection{Asymptotic Matching}

In mathematical terms, our equations possess an \lq\lq internal
boundary layer"~\cite{Bender:78}. The reaction front, defined by $|x +
2\nu\sqrt{t}| = O(1)$, acts as the \lq\lq inner region", while the
diffusion layer, defined by $\sqrt{t} = O(|x + 2\nu\sqrt{t}|)$, acts as the
\lq\lq outer region". For consistency, the \lq\lq inner limit"
($\zeta\rightarrow 0$) of the outer approximation,
Eq.~(\ref{eq:Adef}), 
must match the \lq\lq outer limit" ($\eta
\rightarrow \infty$) of the inner approximation, Eq.~(\ref{eq:Acdef}).
We have shown that $\gamma > 0$ is required to describe the
experimental data, which means that $a(x,t)$ approaches zero uniformly
in the reaction front. Therefore, by matching at zeroth order we
obtain $A^\prime(0) = A_o = 0$, but this does not provide the missing
boundary condition for the reaction front. At the next (linear) order
we have
\begin{equation}
\frac{\partial a}{\partial x}  =  \left\{ \begin{array}{ll}
\frac{\partial A}{\partial \zeta} \frac{\partial \zeta}{\partial x} \sim
\frac{A^\prime(\zeta)}{2\sqrt{t}}, & \ \  \mbox{as} \ t \rightarrow
\infty \ \mbox{with} \ 0 < \zeta < \infty \ \mbox{fixed}  \\
        \frac{1}{\sqrt{t}} \frac{\partial \Ac}{\partial \eta}
\frac{\partial \eta}{\partial x} \sim
\frac{\Ac^\prime(\eta)}{\sqrt{t}}, & \ \  \mbox{as} \ t \rightarrow
\infty, \ \mbox{with} \ |\eta| < \infty \ \mbox{fixed} \end{array} \right.
\end{equation}
and by matching we conclude $\Ac^{\prime}(\infty) = \Ac_1$, where
$\Ac_1(\nu) \equiv A^\prime(0)/2$ can be expressed in terms of
$\nu(q)$ using Eq.~(\ref{eq:A1}). In light of Eq.~(\ref{eq:Bzero}),
the matching condition for $b(x,t)$ is $\Bc(\infty) = 0$.

\begin{figure}
\vspace*{1.5cm}
\epsfxsize= 12.0cm
\centerline{\epsfbox{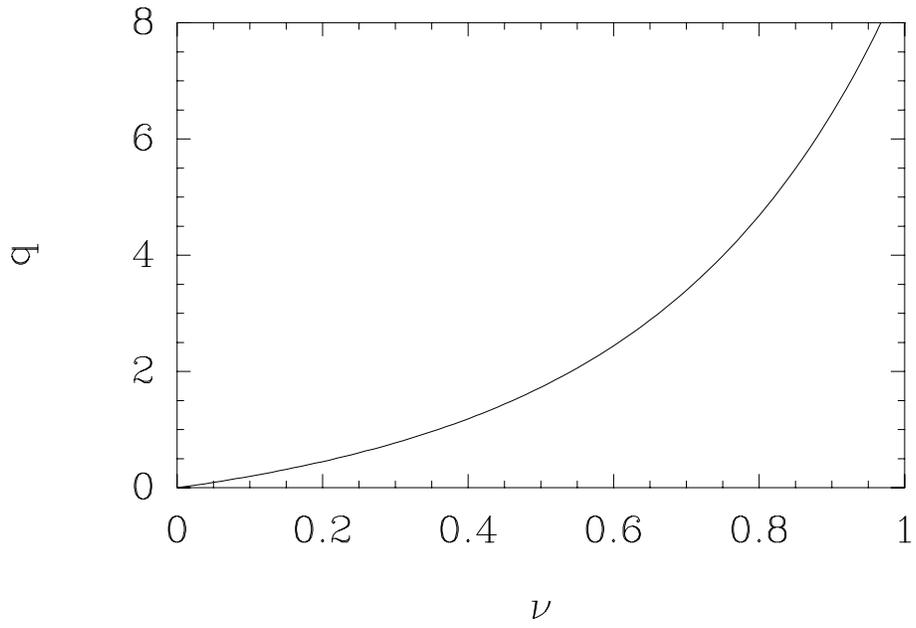}}
\vspace*{0.5cm}
\smallskip
  \caption{The exact asymptotic dependence of $\nu$, the square root
  of the dimensionless diffusion constant of the reaction front, on
  the asymmetry parameter $q$, predicted by Eq.~(\ref{eq:qnu}).  }
\label{fig:nuq}
\end{figure}

The matching conditions allow us to derive an exact expression for
$\nu(q)$ and hence the asymptotic front position $x_f(t) =
2\nu\sqrt{t}$. Taking the limit $\eta \rightarrow \infty$ in
Eq.~(\ref{eq:Acint}) using $\Ac^\prime(\infty) = \Ac_1$ and
$\Bc(\infty) = 0$, we obtain $q \Ac_1(\nu) = \nu$. By substituting
$\Ac_1(\nu)$ from Eq.~(\ref{eq:A1}) we obtain the desired
expression for $\nu(q)$,
\begin{equation}
\nu = F^{-1}(q),
\ \ \ \mbox{where} \ \ \ F(x) \equiv \sqrt{\pi} x e^{x^2}\left[ 1
+ \erf(x) \right] ,\label{eq:qnu}
\end{equation}
which has also been derived by Koza~\cite{Koza:97}. The relation $q =
F(\nu)$ is plotted in Fig.~\ref{fig:nuq} and will be used in the next
section to estimate $q$ from the experimentally measured value of
$\nu$.

With these results, we are led to a second-order, nonlinear
boundary-value problem for the reaction front concentration of the
diffusing species:
\begin{equation}
\Ac^{\prime\prime} = \Ac \ f(1 - \Ac^\prime/\Ac_1), \ \ \ \Ac(-\infty) =
0, \ \ \ \Ac^\prime(\infty) = \Ac_1 .
 \label{eq:Aceq}
\end{equation}
Note that Eq.~(\ref{eq:Aceq}) is invariant to translation $\eta
\mapsto \eta + \eta_o$, where $\eta_o$ is an undetermined 
constant depending on the
initial conditions that precisely defines the location of the reaction
front ({\it e.g.} as the point of maximal reaction rate).  

Since it is
difficult to accurately measure the reaction-front concentration
fields in our experiments, we stop here and refer the reader to the
article of Bazant and Stone~\cite{Bazant:99} for the integration of
this boundary-value problem and other analytic results in the case
$f(b) = b^m$, $m \geq 1$.

\section{Experimental Test of the Theoretical Model}

\subsection{Check of the exact asymptotic predictions}

In section III we showed that as corrosion proceeds the reaction front
moves with the time as $X_f(T) \sim T^{1/2}$ and does not spread
($w(T) \sim T^\alpha$ with $\alpha=0$) and the width $W_d$ of the
depletion layer increases with the time as $W_d(T) \sim T^{1/2}$. In
section IV we showed that these observations are consistent with the
predictions of a one-dimensional $\mbox{A} + \mbox{B\ (static)}
\rightarrow \mbox{C\ (inert)}$
mean-field model with a reaction rate that is first order in the
diffusing species A.
By solving the mean-field equations, we derived not
only the scaling exponents for $X_f(T)$ and $W(T)$ but also the
prefactors and the exact asymptotic shape of the concentration profile
of the diffusing reactant as a function of the reduced coordinate
$\zeta = \frac{X - X_f(T)}{2\sqrt{DT}}$. In this section, we
quantitatively test these theoretical predictions against the
experimental results.

\subsubsection{Movement of the front}

In dimensional units Eq.~(\ref{eq:defxf}) reads:
\begin{equation}
-X_f=2\nu(q)\sqrt{DT}.
\label{eq:Xf}
\end{equation}
Therefore, from a log-log plot of $X_f$ as a function of $T$ one gets
the value of $\nu$, and $q$ can then be deduced from
Eq.~(\ref{eq:qnu}). In our experimental system, $q$ is linearly
related to a characteristic property of the electrolyte, namely the
transference number of the cation, through $q=1-t_+$. To derive the
values of $q$ and $t_+$ from Eq.~(\ref{eq:Xf}), we need an accurate
value of the diffusion coefficient of the electrolyte. $D$ is likely
to depend on the concentration of CuCl$_2$, but to our knowledge, has
not been tabulated for CuCl$_2$. Hereafter, we use the value
$D=(1.0 \pm 0.1) 10^{-5}$cm$^2\cdot$s$^{-1}$, determined independently by our
interferometric technique.

The two sets of experimental data in of Fig.~\ref{fig:front} give
$2\nu\sqrt{D}=(1.7\pm0.1)\ 10^{-3}$cm$\cdot$s$^{-1/2}$, therefore
$\nu=0.27\pm0.02$ and $t_+=0.33\pm0.05$ from Eq.~(\ref{eq:qnu}), $q =
F(\nu)$. Note that $t_+ \approx 0.3$ (for a 0.5~mol.l$^{-1}$
electrolyte) is quite consistent with the corresponding value at
infinite dilution $t_+^{\infty} =0.4$ since $t_+$ is likely to be a
decreasing function of the concentration
\cite{Fleury:91}. 
Although we have not directly measured the transference number $t^+$
of the Cu$^{2+}$ cation, its reasonable value just inferred from the
observed front speed via Eq.~(\ref{eq:qnu}) constitutes a successful
prediction of the one-dimensional mean-field model.

\subsubsection{Width of the depletion zone and whole concentration profile}

\begin{figure}
\vspace*{1.5cm}
\epsfxsize= 12.0cm
\centerline{\epsfbox{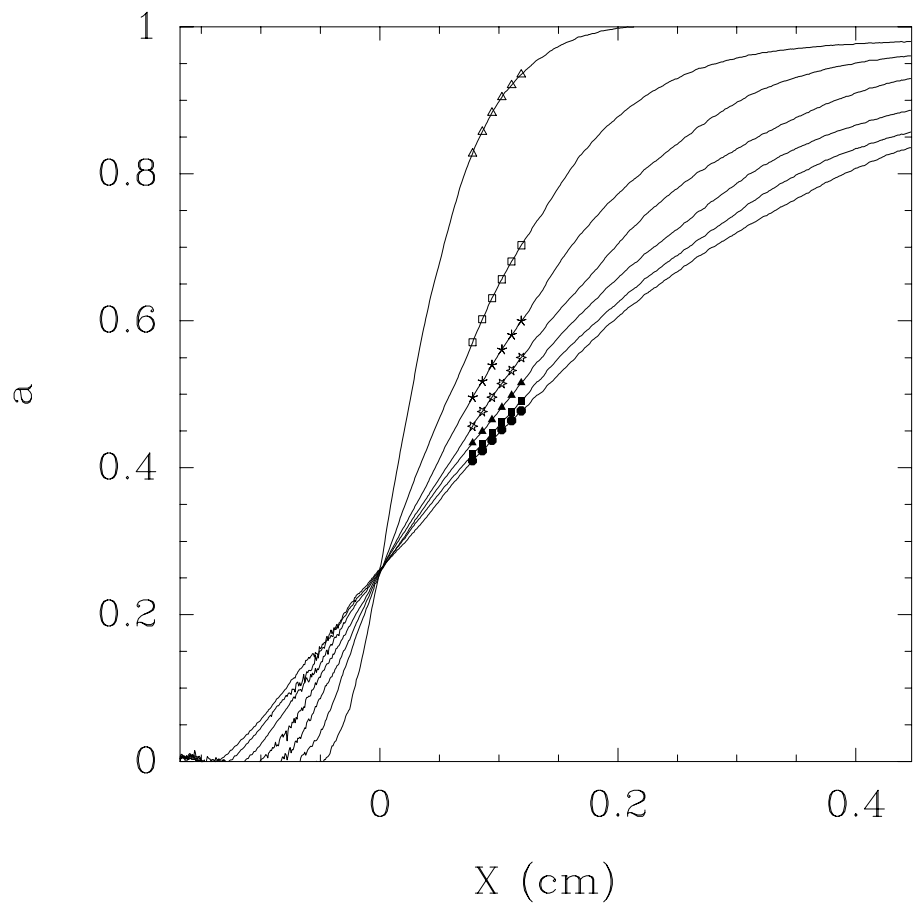}}
\vspace*{0.5cm}
\smallskip
  \caption{One-dimensional concentration profiles extracted from the
    two-dimensional data. The deposit has been grown from a
    1~mol.l$^{-1}$ CuCl$_2$ solution at $j=68$~mA.cm$^{-2}$  during 15
    minutes. The concentration profiles are shown every 15 minutes
    after the current had been switched off. The different symbols are
    added on each profile to differentiate the recording times. These
    symbols will be used on the next representations of the
    concentration profiles in Figs 9 and 10.
}
\label{fig:profils1}
\end{figure}

\begin{figure}
\vspace*{1.5cm}
\epsfxsize= 12.0cm
\centerline{\epsfbox{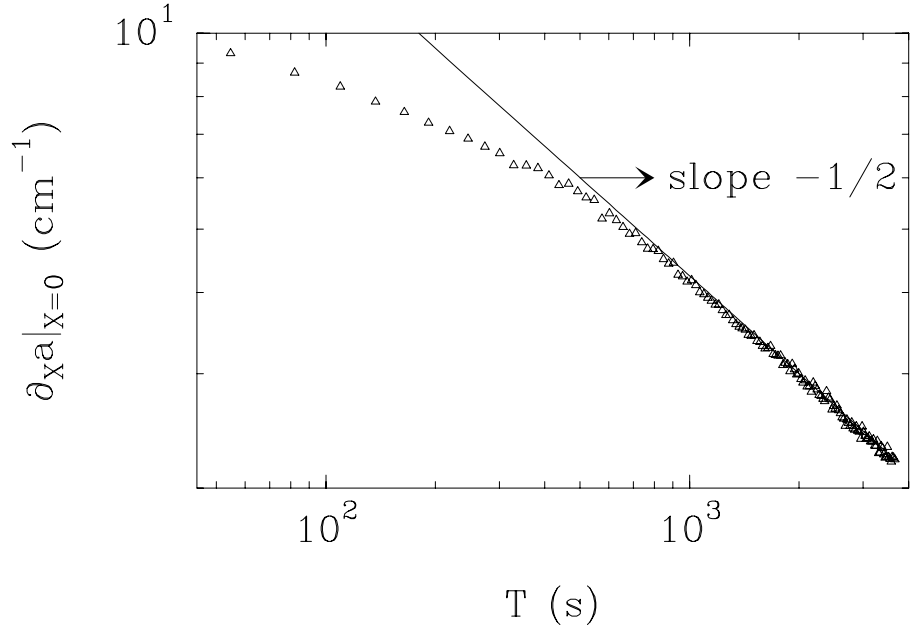}}
\vspace*{0.5cm}
\smallskip
  \caption{Log-log plot of the temporal evolution of the derivative of
  $a(X,T)$ at $X=0$ as a function of $T$. Same parameters as in
  Fig.~\ref{fig:profils1}. The plain line corresponds to the
  prediction of Eq.~(\ref{eq:grad}) with $D=10^{-5}$~cm$^2$.s$^{-1}$ and
  $q=0.79$.  }
\label{fig:gradient}
\end{figure}

In this section, we analyze the experiments  performed with a
higher electrolyte concentration, namely 
1.0~mol.l$^{-1}$ CuCl$_2$.  The concentration profile in the 
laboratory frame can be written in dimensional units using
Eq.~(\ref{eq:diffA}) and the definition of $\zeta$~:
\begin{equation}
a(X,T)=\frac{\erf(X/2\sqrt{DT}) + \erf(\nu)}{1 +\erf(\nu)}.
\label{eq:labo}
\end{equation}
Note that 
$a(X,T)$ is used in the experimental parts to 
denote $\rho_A(X,T)/\rho_A^o$.
A characteristic feature of these profiles (and the experimental data
in Fig.~\ref{fig:profils1}) is that they exhibit a
fixed point with ordinate:
\begin{equation}
a(X=0,T)=\frac{\erf(\nu)}{1+\erf(\nu)} \ \ . \label{eq:afixed}
\end{equation}
Since $a(0,T)$ depends only on $q$, a value of $q$ can be deduced from
Fig.~\ref{fig:profils1}, which shows the concentration profiles during
the corrosion of a copper deposit obtained by electrodeposition from a
1.0~mol.l$^{-1}$ CuCl$_2$ solution. We find $a(X=0,T)=0.25\pm0.01$
which implies $\nu= 0.30 \pm0.01$.
From Eq.~(\ref{eq:qnu}) the mean-field model would predict $q =
0.79\pm0.06$. As expected, the inferred value of the transference
number, $t_+=1 - q= 0.21\pm0.06$, for this 1.0 ~mol.l$^{-1}$ CuCl$_2$
solution is lower than the value of $0.33\pm0.05$ at 0.5~mol.l$^{-1}$
computed above.
This value is somewhat smaller than expected based on concentration
effects (see below).
Note that we have not directly measured the ratio
$q=\rho_A^o/\rho_B^o$ 
or the transference number $t^+$
in the experiments described in this paper, but the
value of $q=0.79$ just obtained from Eq.~(\ref{eq:afixed}) is
necessary for comparison with the mean-field model (without any other
adjustable parameters). Therefore, we will use $q=0.79$ in the
following analysis of the experimental runs in 1~mol.l$^{-1}$ CuCl$_2$
electrolyte.

>From Eq.~(\ref{eq:diffA}) the width $W_d$ of the diffusion layer (with 
dimensions) is given by:
\begin{equation}
W_d(T)=\left( \left. \partial_Xa(X,T) \right|_{X=X_f} \right) ^{-1}
=\left( \frac{\exp(-\nu^2)}{\sqrt{\pi DT}(1+\erf(\nu))} \right) ^{-1}.
\end{equation}
From an experimental point of view, it is simpler to measure 
$a(X,T)$ at $X=0$ rather than at $X=X_f(T)$, so we consider  
the temporal evolution of the gradient of $a(X,T)$ at
$X=0$. From Eq.~(\ref{eq:labo}) we obtain:
\begin{equation}
\left.\frac{\partial a(X,T)}{\partial
X}\right|_{X=0}=\frac{1}{\sqrt{\pi DT}(1+\erf(\nu))}
\label{eq:grad}
\end{equation}
and $W_d(T)=\exp(\nu^2)/\left. \partial_Xa(X,T) \right|_{X=0}$.
Figure~\ref{fig:gradient} shows the quantitative agreement between the
experimental values of $\left.\partial_Xa\right|_{X=0}$ and the
function of Eq.~(\ref{eq:grad}) plotted for $D=10^{-5}$cm$^2\cdot$s$^{-1}$ and
$q=0.79$. Note that $D$ and $q$ are deduced from previous analysis and
are not adjustable parameters.

\begin{figure}
\vspace*{1.5cm}
\epsfxsize= 12.0cm
\centerline{\epsfbox{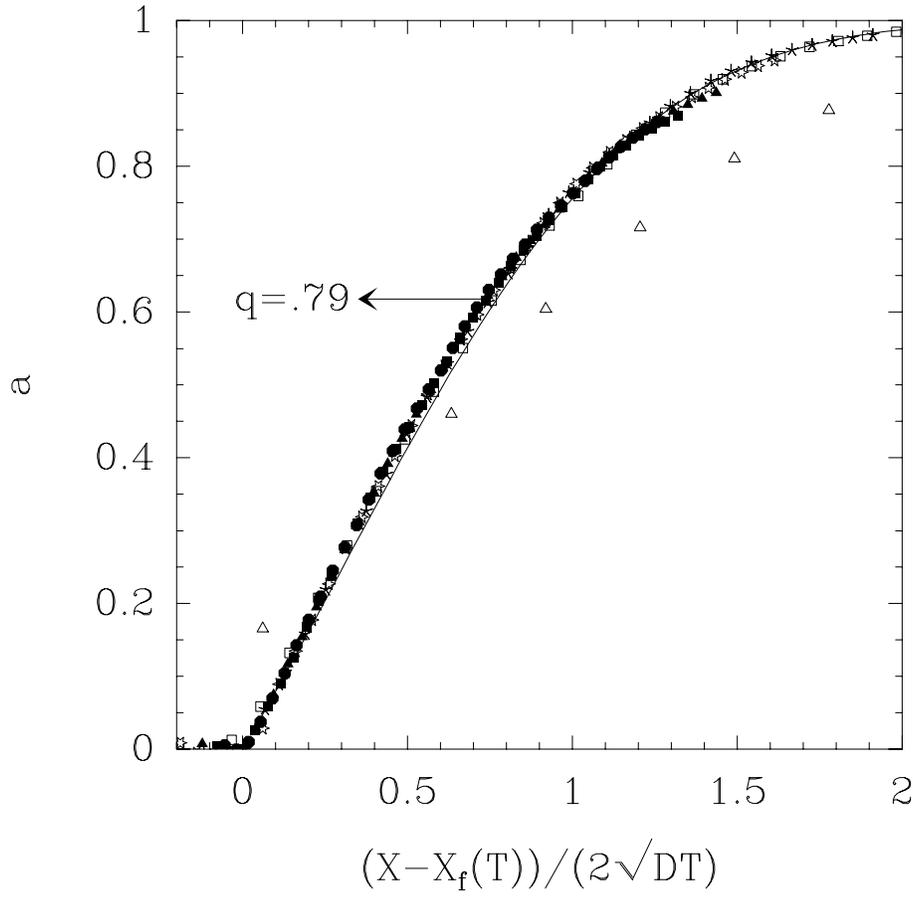}}
\vspace*{0.5cm}
\smallskip
  \caption{
Collapse of the experimental concentration data in the diffusion layer
plotted versus the similarity variable $(X-X_f)/2\protect\sqrt{DT}$
compared with the theoretically predicted asymptotic experimental
similarity function $A(\zeta)$ (the solid line). The profiles are the
same as those plotted in Fig.~\ref{fig:profils1}, but only one point
out of 20 is shown on this plot for clarity.
}
\label{fig:profils2}
\end{figure}

Continuing our quantitative analysis of the experimental
concentration  field, we plot in Fig.~\ref{fig:profils2} the
asymptotic  shape of the concentration profile. To determine
$a(\zeta)$ from  $a(X,T)$, we compute $\zeta$ using
$\zeta=\frac{X}{2\sqrt{DT}}+\nu(q)$, with $q=0.79$ and 
$D=10^{-5}$cm$^2$.s$^{-1}$ and adjust the origin of the abscissa
to the initial front of copper position,
to ensure that $\At(\zeta=0,T)=0$ for all $T$.
For comparison we also show in the same plot the theoretically
predicted function $A(\zeta)$ function computed from
Eq.~(\ref{eq:diffA}) and (\ref{eq:qnu}) with $q=0.79$.

\begin{figure}
\vspace*{1.5cm}
\epsfxsize= 12.0cm
\centerline{\epsfbox{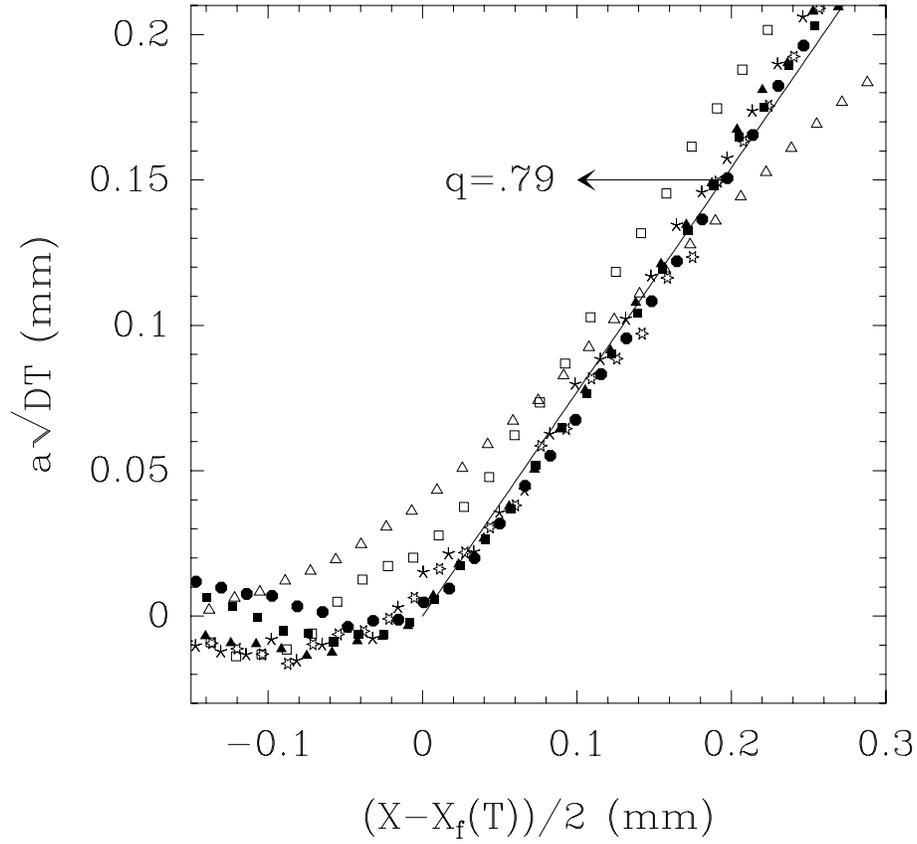}}
\vspace*{0.5cm}
\smallskip
  \caption{
Collapse of the experimental concentration data in the reaction front
plotted versus the similarity variable $(X-X_f)/2$. The solid line
shows the linearized extension of the similarity function $A(\zeta)$
from the diffusion layer (see Fig.~\ref{fig:profils2}) extended into
the reaction front. These profiles are the same as those plotted in
Fig.~\ref{fig:profils1}, but only one point out of 4 is shown on this
plot for clarity. The negative concentration values are artifacts of
the interferometric technique and have no physical meaning.
}
\label{fig:profils3}
\end{figure}

To focus on the region of the reaction front,
the experimental data is plotted in fig~\ref{fig:profils3} according
to the linearized version of Eq.~(\ref{eq:diffA})
\begin{equation}
a\sqrt{DT}= A^\prime(0) \frac{X-X_f}{2} 
=\frac{2 e^{-\nu^2}} {\sqrt{\pi}(1+\erf(\nu))}\frac{X-X_f}{2} \; ,
\end{equation}
Since $(X- X_f)/2$ is proportional to the reaction-front similarity
variable $\eta$ in Eq.~(\ref{eq:etadef}), the mean-field model would
predict a collapse of this data to a single curve given by the solution
of Eq.~(\ref{eq:Aceq}).

Unfortunately the noise in the experimental data washes out the exact
concentration profiles in the reaction front on this scale, but it is
clear that the width of the reaction front has the asymptotic scaling $w
\sim t^\alpha$ with $\alpha = 0$. Moreover, the asymptotic shape of
the concentration distribution is quite consistent with the solutions
to Eq.~(\ref{eq:Aceq}) given in Ref.~\cite{Bazant:99}. Note that the
decay of slope the reaction-front concentration $\Ac^\prime(\eta)$
toward its limiting in the diffusion layer $\Ac_1$ in
Fig.~\ref{fig:profils3} appears to be quite fast. If this decay were
exponential rather than a (much slower) power law, then according to
the mean-field model~\cite{Bazant:99} the reaction rate would have to
be first order in the static reactant $m=1$, i.e. $f(b) =b$ or $r(a,b)
= ab$, but it is impossible to reach this conclusion
definitively from our data.

As shown in Figs~\ref{fig:profils2} and \ref{fig:profils3}, all of the
measured concentration profiles collapse to the single asymptotic curve
predicted for $q=0.79$ over the whole length scales investigated in
the experiment. This quantitative agreement between the experimental and
theoretical concentration profiles of the diffusing reactant  
independent of the length scale
strongly support our modeling of this corrosion experiment with 
a one-dimensional 
$\mbox{A} + \mbox{B\ (static)}
\rightarrow \mbox{C\ (inert)}$ 
 mean-field model.

\subsection{The transient}

The $\mbox{A}+\mbox{B}\rightarrow \mbox{C}$ mean-field model with two
diffusing reactants exhibits many surprising and
nontrivial behaviors at short times (see
\cite{Taitelbaum:98} and references therein, \cite{Yen:96},
\cite{Chopard:93}).  In this case, some microscopic parameters like
the reaction constant(s) can be determined from these short time
behaviors. In particular, at 
a time inversely proportional to the microscopic reaction constant, the
global reaction rate switches from an initial $t^{1/2}$ increase to a
subsequent $t^{-1/2}$ decrease
\cite{Taitelbaum:98}. Moreover, in the reversible
$\mbox{A}+\mbox{B}\rightleftharpoons \mbox{C}$ system, a crossover
between irreversible and reversible regimes can be observed at long
times \cite{Chopard:93} and the value of the backward reaction
constant can be inferred from the crossover time \cite{Yen:96}.

In the present case of one static reactant, it is also possible to
express the transient decay to the asymptotic solution in terms of the
reaction orders $m$ and $n$ for the one-dimensional mean-field
model~\cite{Bazant:99}.  In our experiment, however, the transient
behavior is determined by a superposition of different mechanisms
since our system is not really one\--dimensional or homogeneous. We
now show that the transient behavior appears to be governed by
two-dimensional geometric effects that hide the kinetic features by
analyzing the detail the experimental runs.

\begin{figure}
\vspace*{1.5cm}
\epsfxsize= 12.0cm
\centerline{\epsfbox{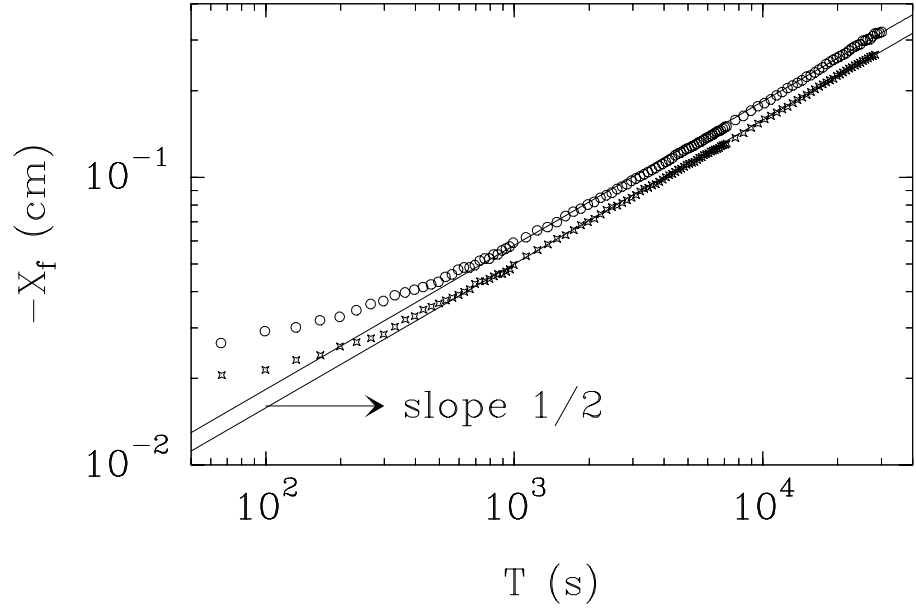}}
\vspace*{0.5cm}
\smallskip
  \caption{Log-log plot of the position of the
  reaction front $X_f$ as a function of
    time $T$ for two different experimental
    runs in CuCl$_2$ 0.5~mol.l$^{-1}$ for deposits
    grown at $j=40$~mA.cm$^{-2}$. The solid lines
    of slope $\frac{1}{2}$ represent the predictions of the one-dimensional,
    mean-field theory, given by Eq.~(\ref{eq:Xf}), with
    $D=10^{-5}$~cm$^2$.s$^{-1}$, in the cases $q=0.6$ and $q=0.73$. }
\label{fig:front}
\end{figure}

Looking at Fig.~\ref{fig:pattern}(a), note that the concentration
field is not one-dimensional at the early stages of the corrosion
experiment: the isoconcentration lines closely follow the jagged
outline of the deposit in the region near the tips.  
The amplitude $G$ of the modulation of the leftmost isoconcentration line
(the closest to the copper cluster) is about 0.4~mm.
This system clearly cannot be viewed as
one-dimensional until the front has traveled at least a distance on
the order of $G$.  In Fig.~\ref{fig:front}, note that the time
of the transient regime (before the asymptotic $t^{1/2}$ behavior sets
in) closely corresponds to the time needed for the front to move
across a distance $G \sim 0.4mm$. (This two-dimensional geometric
effect also may explain why the initial movement of the front is
slower than the asymptotic behavior, as shown in Fig.~\ref{fig:front}.)

\begin{figure}
\vspace*{1.5cm}
\epsfxsize= 12.0cm
\centerline{\epsfbox{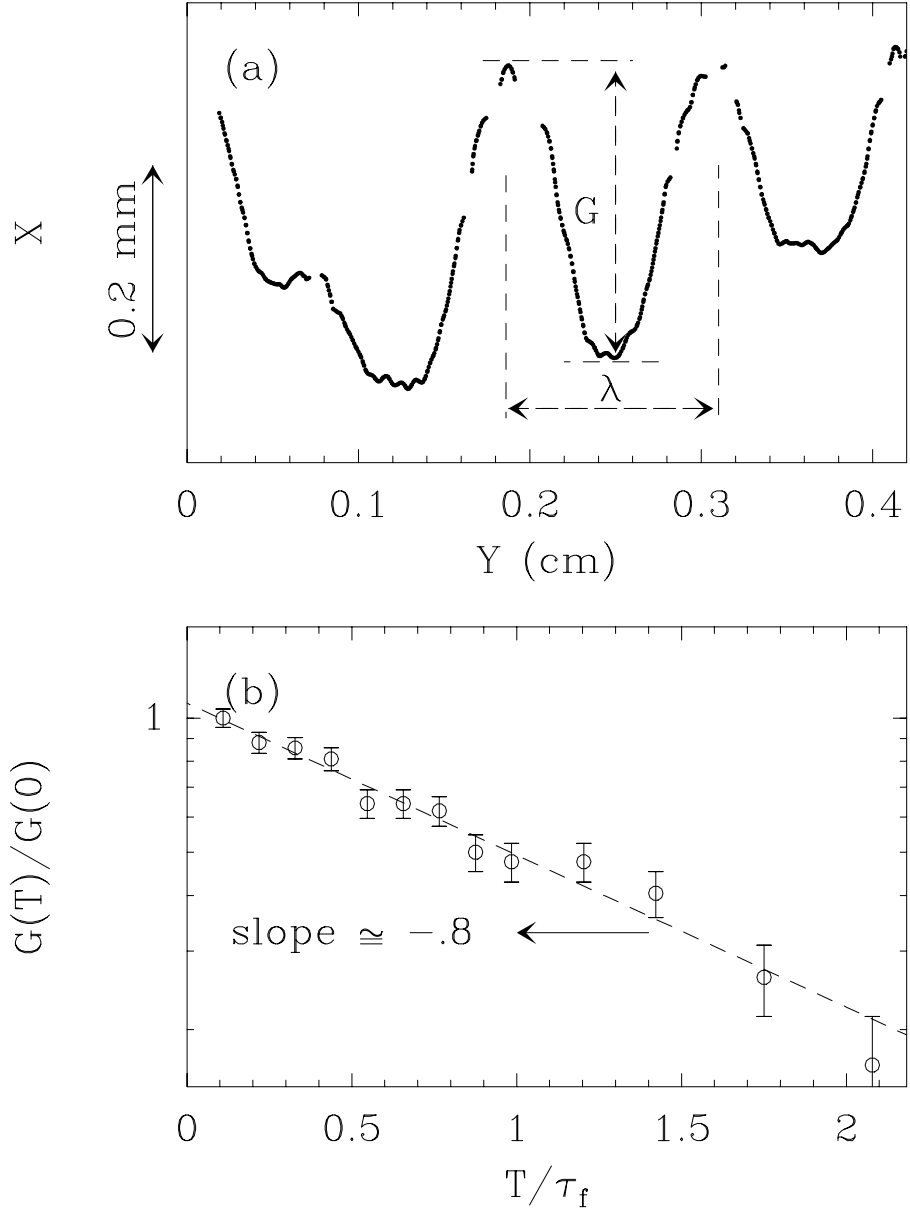}}
\vspace*{0.5cm}
\smallskip
  \caption{Relaxation of the two dimensional initial concentration
field at the beginning of the dissolution.
(a) Isoconcentration line $a=0.1$, for $T= 27$~s.
The deposit has been grown from a 1~mol.l$^{-1}$ CuCl$_2$  solution, at
$j=68 $mA.cm$^{-2}$ during 15 minutes.
(b) Log-linear plot of the evolution of the amplitude $G$ of the
modulation of $A$ concentration, as shown in (a), versus
the reduced time $T/\tau_f = 4\nu^2DT/G(0)^2 $.}
\label{fig:relax}
\end{figure}

To further support this hypothesis, we now study the relaxation
dynamics of the concentration field.  In Fig.~\ref{fig:relax}-(a), is
plotted the isoconcentration line corresponding to
$a=\rho_a/\rho_a^o=0.1$, just after the current has been switched
off. This line is not continuous, because the concentration field
cannot be extracted by interferometry in the zones containing the
deposit. This line defines a function $X(Y)$ roughly periodic, of
amplitude $G(T)$ and period $\lambda \sim 1$mm.  It is
reasonable to expect that the characteristic time for the relaxation
of this modulation of the concentration field toward a flat two
dimensional profile
is the time $\tau_{f}$ needed for the front of copper to move from its
starting position ($X_f (T=0)$) on the length scale $G(T=0) =
0.3$~mm$= 2 \nu
\sqrt{D\tau_{f}}$, which yields the estimate
$\tau_{f}=G(0)^2/4\nu^2D \sim 250$s. 
Moreover, in light of the analysis of Krug~\cite{Krug:91} described
below, it is also reasonable to expect that the functional form of the
decay will be exponential.

In Fig~\ref{fig:relax}-(b), we plot $\log(G(T)/G(0))$ as a
function of the dimensionless time $T/\tau_{f}$. The relaxation is
well fitted by an exponential function, with a characteristic time
close to $\tau_{f}$, which supports our hypothesis.  Therefore in our
experiments, the transient behavior is directly linked to the
relaxation of the initial two dimensional concentration field towards
a Y-invariant profile and cannot provide information on the kinetics
independently.

\subsection{Physical relevance of the one\--dimensional mean\--field model}

In the previous sections, we have demonstrated the quantitative
agreement between the behavior of our thin\--gap corrosion system and
various predictions of a one-dimensional mean field model. This
agreement is not obvious {\em a priori}, and therefore we close in
this section by giving physical arguments to explain this surprising
fact.

\subsubsection{No inhibition of diffusion or reaction by CuCl}

In Fig.~\ref{fig:pattern} we see that the product of the reaction does
not seem to disturb the concentration field of the diffusing reactant
A.  To understand this fact, we 
consider the volume occupied by the product
CuCl in the cell. 
We know from Eq.~(\ref{eq:q}) that the mean concentration of copper
before the dissolution is $\approx 2\rho_A^o$. We deduce from
Eq.~~(\ref{eq:bilan}) that if CuCl does not diffuse (which is verified
in our experiments), the mean concentration of CuCl is twice the
initial concentration of copper, which is approximately four times the
initial concentration of CuCl$_2$ in the bulk, {\it i.e.}  
2~mol.l$^{-1}$.  Since the density and molecular weight of CuCl are 
3.38~g.cm$^{-3}$ and 99~g.mol$^{-1}$ respectively, the volume occupied by
the solid CuCl after the dissolution is roughly 5\% of the
total local volume.
Therefore, the small crystals of CuCl do not significantly alter the
volume free for the diffusion of CuCl$_2$.  Moreover, because the CuCl
crystalites do not adhere to the copper metal branches and fall to
the bottom of the thin gap cell, the surface of the copper cluster is
constantly renewed and ``ready'' for corrosion by CuCl$_2$.

\subsubsection{Stable front, asymptotically one\--dimensional}

The fact that the dissolution process builds a stable (flat) interface
can be understood by considering that diffusion-limited corrosion is
the ``time-reversed'' process of diffusion-limited aggregation and
that the fluctuations of the interface decay rather than grow to
reach a stable flat front asymptotically.  Krug~\cite{Krug:91} showed
that periodic perturbations of a flat front of wavelength $\lambda$ in
the direction perpendicular to the direction of motion of the
interface would decay with a characteristic time $\tau=\lambda/v$. The
stability of the corrosion front can therefore be qualitatively
understood with the following argument: the electrolyte 
most easily reaches 
the most exposed or least screened parts of the copper deposit.  These
bulges are dissolved first, and the interface is smoothed.

\subsubsection{Relevance of 1D approximation of the concentration field}

In the long-time asymptotic regime, the modulation of the initial
concentration of reactant A (CuCl$_2$) relaxes toward a flat
concentration profile along the direction $Y$ whose shape is given by
Eq.~(\ref{eq:diffA}). However, the concentration of the static
reactant B (Cu), as well as the concentration of the product C
(CuCl) keep a periodic shape along the $Y$ direction, which somehow
does not alter the one-dimensional asymptotic solution.  The largest
characteristic length of the deposit in the direction parallel to the
front ($Y$) is the mean distance $\lambda$ between the trees. This puzzling
observation can be understood by comparing the relaxation time of the
perturbations of $\rho_A(X,Y)$ along $Y$, $\tau_{d}
\sim \lambda ^2/D$, with the time needed by the front of copper to move
on the same length, $\tau_{f} \sim \lambda/\dot{X_f} = 
\lambda \sqrt{T}/\nu\sqrt{D}$. Since $\tau_{f}$ increases with time
$T$, in the asymptotic regime it will be much greater than $\tau_d$.
Therefore, whereas $\rho_B$ is highly correlated along the $Y$
direction due to the structure of the solid deposit, there are
eventually no fluctuations in $\rho_A$ along this direction.

\subsubsection{Departure from pure diffusion in the reaction zone}

The fact that the transference number $t^+$ deduced from $t^+ = 1 - q$
and the inferred value of $q = F(\nu)$ from Eq.~(\ref{eq:qnu})
decreases significantly from 0.33 to 0.21 when the concentration of
CuCl$_2$ is increased from 0.5~mol.l$^{-1}$ to 1~mol.l$^{-1}$ is
unlikely to be caused solely by a pure salt-concentration effect. It
is also possible that convection produced by the sedimentation of CuCl
crystallites toward the bottom of the cell could artificially increase
the effective diffusion coefficient close to the reaction front by
convective mixing.  This would cause an increase of $\nu(q)$ (the
prefactor for the speed of the reaction front) which could at least
partly explain the difference in the inferred $q$ values,
and therefore also in the effective $t^+$ values.

\section*{Conclusion}

We have shown that after long times the corrosion of highly porous
copper clusters can be understood as a one-dimensional, homogeneous,
mean-field $\mbox{A}+\mbox{B} \rightarrow \mbox{C}$ reaction-diffusion
process with one diffusing and one static reactant.  This is the first
experimental analysis of such a situation where
only one reactant is free to
diffuse through the other one.  Whereas one would expect highly
complex dynamics and a possible breakdown of the mean-field
approximation when the reaction is confined to a porous (fractal)
interface, we show that in this particular corrosion system, the
dynamics are equivalent to those expected for an homogeneous
system. The strength of our demonstration is built on
precise measurements of the concentration field of the diffusing
species by interferometry which are compared quantitatively with
analytical predictions of the one-dimensional mean-field model.

\acknowledgements

We are very grateful to Y. Sorin and G. Gadret for their technical
assistance with the optical set\-up. We also thank H. A. Stone, E.
Cl\'ement, J. Elezgaray and A. Arneodo for stimulating and fruitful
discussions. This work was supported (mainly) by the Centre National
d'Etudes Spatiales under Grant n$^{\circ}$97/CNES/071/6850 and also
(partially) by an NFS infrastructure grant to the MIT Department of
Mathematics.


%
%


\end{document}